\documentclass[a4paper]{jpconf}
\usepackage{amssymb}
\usepackage{graphicx}
\usepackage{amsthm}
\usepackage{latexsym}
\newcommand{\eprint}[1]{preprint [arXiv: #1]}

\newcommand{\Bgdist}{Naive spatial distance}
\newcommand\beq{\begin{equation}}
\newcommand\eeq{\end{equation}}
\newcommand\bea{\begin{eqnarray}}
\newcommand\eea {\end{eqnarray}}

\newcommand{\ie}{\emph{i.e.},\,}

\def\un\a{{\underline\alpha}}

\def\a{{\alpha}}

\newcommand{\bgdist}{naive spatial distance}

\begin{document}
\title{\textbf{Emergent Continuum Spacetime from a Random, Discrete, Partial Order}\footnote{Based on talk given by P. Wallden at the NEB XIII conference}}

\author{\textbf{David Rideout}}
\address{Perimeter Institute for Theoretical Physics \\ 31
Caroline Street North, Waterloo, Ontario N2L 2Y5, Canada}

\ead{\textbf{drideout@perimeterinstitute.ca }}

\author{\textbf{Petros Wallden}}
\address{Raman Research Institute, Theoretical Physics Group \\ Sadashivanagar, Bangalore - 560
080, India}

\ead{\textbf{petros.wallden@gmail.com}}

\date{}

 \vspace{1cm}

\begin{abstract}

There are several indications (from different approaches) that
Spacetime at the Plank Scale could be discrete. One approach to
Quantum Gravity that takes this most seriously is the Causal Sets
Approach. In this approach spacetime is fundamentally a discrete,
random, partially ordered set (where the partial order is the causal
relation). In this contribution, we examine how timelike and
spacelike distances arise from a causal set (in the case that the
causal set is approximated by Minkowski spacetime), and how one can
use this to obtain geometrical information (such as lengths of
curves) for the general case, where the causal set could be
approximated by some curved spacetime.
\end{abstract}

\section{Motivation}

One of the main problems that any discrete/combinatoric approach to
Quantum Gravity\footnote{Loop quantum gravity, spin foams, causal
dynamical triangulations, causal sets, etc.} faces is the so called
\emph{inverse problem} \cite{inverse_problem}, which has two sides:
(a) How do smooth continuum-like structures emerge from the
underlying discrete dynamics and (b) how would we recognize that our
discrete entity happens to be something that is smooth and
continuum-like. Here we will focus on the second part of the inverse
problem, for the specific approach to quantum gravity known as
\emph{causal sets} (see Ref. \cite{blms} or  Refs.
\cite{causet_reviews,Dowker:2006} for reviews).
\section{This paper}

This contribution is largely based on Ref. \cite{diameters}, where
detailed numerics and other related work can be found. In Section
\ref{causets} we introduce the causal sets approach to Quantum
Gravity, in Section \ref{previous attempts} we review previous
attempts to recover continuum properties from a causal set, in
Section \ref{section spatial distance} we introduce the new
suggestion to recover spatial distance in a ``flat'' causal set and
in Section \ref{section curved spacetime} we mention how this can be
used %to generalize
for a general (curved spacetime) causal set. We
summarize and conclude in Section \ref{conclusions}.

\section{Causal Sets}\label{causets}

Causal sets is a fundamentally discrete approach to quantum gravity.
There are several indications that spacetime at the Plank scale
could be discrete. These involve the finite black hole entropy, the
infinities in general relativity and quantum field theory, as well
as indications from other approaches (the discrete volume spectrum in
loop quantum gravity, effective minimum length due to dualities in
string theory, and others). Causal sets take these indications
literary and start with a spacetime that is not a differentiable
manifold, but rather a locally finite partially ordered set. More
specifically, a causal set is:

\begin{enumerate}

\item  Partially Ordered Set: a set $\mathcal P$ with
relation $\prec$, such that $\forall\quad x,y,z\in\mathcal
P:$\begin{itemize}\item[(a)]$ x\prec y$ and $ y\prec z \Rightarrow
x\prec z$: Transitivity \item[(b)] $ x\nprec x$: Irreflexive (with (a) this forbids closed loops)\end{itemize}

\item Locally Finite: $\{z|x\prec z\prec y\}$ is a finite
set for all $ x,y\in\mathcal P$.

\item The elements of the set correspond to spacetime points, the order relation ($\prec$) is the causal relation
between spacetime points, and the number of elements corresponds to
the spacetime volume.

\end{enumerate}
The second condition is what makes the causal set approach a
discrete approach to quantum gravity, since it guarantees that in a
finite volume there will be only be a finite number of elements.

A theorem due to Malament (see Ref. \cite{Malament}) states that one
can recover the conformal metric of a spacetime (Lorentzian
manifold) using solely the causal (partial) order. This means that
all degrees of freedom are encoded in the causal structure, apart
from a scale factor. However if we assume spacetime to be discrete,
we can fix the scale factor, by counting the number of elements
(corresponding to spacetime volume). This leads us to the central
conjecture of causal sets ``Hauptvermutung'':

\begin{quote}Two distinct, non-isometric spacetimes cannot arise
from a single causal set.
\end{quote}
In order to be precise, we need to define when we say that a causal
set is approximated accurately by a spacetime manifold. For this we
define the concept of \emph{faithful embedding}:

\begin{quote} A faithful embedding is a map $\phi$ from a causal set
$\mathcal P$ to a spacetime $\mathcal M$ that:\begin{enumerate}

\item preserves the causal relation (\ie $x\prec y\iff
\phi(x)\prec \phi(y)$) and \item is ``volume preserving'', meaning
that the number of elements mapped to every spacetime region is
Poisson distributed, with mean the volume of the spacetime region in
fundamental units, and \item $\mathcal M$ does not possess curvature
at scales smaller than that defined by the ``intermolecular
spacing'' of the embedding (discreteness scale).\end{enumerate}
\end{quote}
Thus the central conjecture reads as ``a causal set
cannot be faithfully embedded into two non-isometric spacetimes''.  One
would like to use these causal sets as fundamental entities in order
to construct a quantum theory of gravity. Work on the possible
dynamics and phenomenology has been carried out recently (see Refs.\
\cite{Brightwell:2007,johnston,sverdlov} and Refs.\
\cite{Sorkin:2007,Sorkin:dalembertian} respectively). However, this
contribution deals with some kinematical questions, and in
particular with the question ``how can one derive effective
continuum properties such as spatial distance starting from a causal
set, when this causal set can be approximated by a spacetime''. This
would arguably, among other things, help us prove the central
conjecture.

\begin{figure}[hbtp]
\center
\includegraphics[scale=0.25]{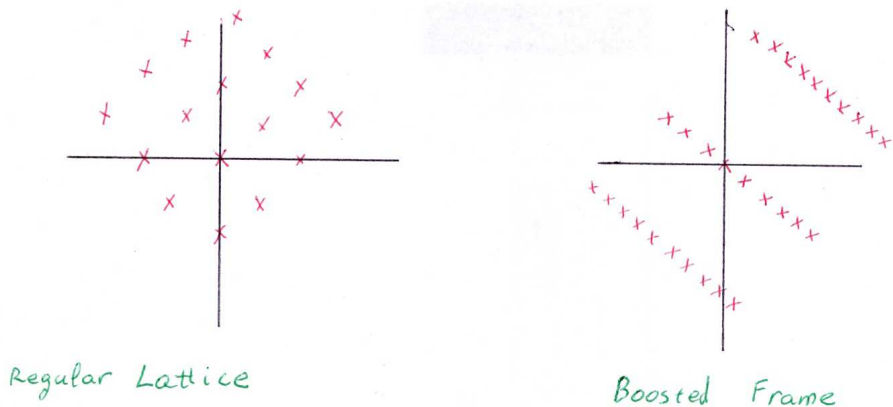}
\caption{\label{lorentz-invariance}Two dimensional Minkowski
spacetime (a) Left: Regular lattice in rest frame (b) Right: Same
lattice in a boosted frame}
\end{figure}

Before moving to the main topic of this contribution, let us make a
remark about Lorentz invariance versus discreteness (see Ref.\
\cite{lorentz_invariance}). It is well known that these two concepts
are not (easily) compatible. To have Lorentz invariance, we would
like to have (approximately) equal number of elements for every
volume of equal size (number-volume correspondence) and this should
hold independently of frames. In particular, let us consider a
regular lattice approximating a 2-dimensional Minkowski spacetime
(see left side of Figure \ref{lorentz-invariance}). The
number-volume correspondence seems to hold, however if we consider
the same in a boosted frame (right side of Figure
\ref{lorentz-invariance}), we can clearly see that there are some
big ``voids'' and the lattice cannot be thought of as being
approximated by Minkowski spacetime.

It can be shown that the unique way\footnote{Subject to some
mathematical subtlety, which allows for some more essentially same
solutions} to achieve discreteness and Lorentz invariance, is by
considering a \emph{Poisson sprinkling}, i.e.\ sprinkle randomly
elements in the spacetime in question, with probability determined
by the poisson distribution with average number of elements
proportional to the size of the spacetime volume of each region.

\beq P(n)=\frac{(\rho V)^n e^{-\rho V}}{n!} \;, \eeq where $\rho$ is
the density, usually fixed at one per Plank volume. Therefore, we
can say that a causal set is a \emph{random}, discrete, partially
ordered set.

\section{Previous Attempts}\label{previous attempts}

Much has been written about causal set kinematics (see Refs.\
\cite{bg,myrheim,thooft,dimension,valdivia,homology}); here we will
focus on a way to achieve timelike and spacelike distances for a
causal set that arises from a sprinkling in a Minkowski spacetime.
Moreover we will consider how (some of) these can be generalized for
curved spacetimes, such as determining the length of curves.

\subsection{Timelike distance}

It can be shown that the concept of ``timelike'' distance (or else
the proper time between two timelike separated elements), can be
easily defined for a ``Minkowski'' causal set and can also be
generalized for a ``curved'' causal set. First let us define a
\emph{link}.  It is the most basic relation between elements, a
relation that cannot be deduced by transitivity: \beq  x\prec y
\textrm{ are linked if } \nexists \quad z|\quad x\prec z\prec y\eeq
A chain is a collection of elements $C$ such that for all $x,y\in
C\quad x\prec y$ or $ y\prec x$. Proper time $d(x,y)$, between two
related elements $x\prec y$, we define to be (following Ref.
\cite{bg}) the number of links $L$ in the \emph{longest}
 chain between (and including) $x$ and $y$\footnote{Note that the Lorentzian character of the partial order is manifested in this definition.
Graphs (of `finite valence'), on the other hand, naturally embed
into Euclidean spaces, and hence one generally defines distance in
terms of \emph{shortest} paths on a graph.}: \beq d(x,y):=L \;.\eeq In
other words, one considers all chains starting from $x$ and ending
at $y$, and  counts the number of links in a largest
one\footnote{Typically there exist more than one biggest chain.}.
This definition is intrinsic to the causal set, and does not depend
on whether it can be faithfully embedded into a manifold, nor on the
expected dimension of such a manifold. In \cite{bg} (and references
therein) it is shown that, in the case of a casual set $\mathcal P$
which arises by a sprinkling of density $\rho$ into $d$-dimensional
Minkowski space (so one gets a faithful embedding $\phi: \mathcal{P}
\to M$), the distance $d(x,y)$ is proportional to the proper time
between the endpoints $\phi(x)$ and $\phi(y)$.  In particular the
authors state that
\beq L(\rho V)^{-1/d}\rightarrow m_d \; \mathrm{ as } \; \rho
V\rightarrow \infty \;, \label{bg_result} \eeq for some constant
$m_d$ which depends upon the dimension. Here $V$ is the spacetime
volume of the causal interval $J^+(x) \cap J^-(y)$ ($J^\pm(x)$
represents the causal future/past of $x$ respectively). The exact
value of $m_2$ is known to be $2$, while for other dimensions some
bounds exist: $1.77\leq m_d\leq 2.62$. Note that this expression
holds at the limit where $\rho V$ goes to infinity, however
numerical analysis for the behavior before the asymptotic limit can
be found in Ref. \cite{diameters}.

\subsection{\Bgdist{}}

In the continuum, for Minkowski spacetime, one can show %it can be shown
the
following: For two spacelike points $x,y$, their distance is identically
the same as the minimum timelike-distance between a point $z\in
J^+(x)\cap J^+(y)$ and a point $w\in J^-(x)\cap J^-(y)$, in other
words the shortest time taken to go from the common past to the
common future. Inspired by this one can define \bgdist{} on a causal
set\footnote{This proposal had been made in the past (e.g.\ Ref.\
\cite{bg}), but was rejected by the original authors themselves, for
 essentially the same reasons that we present below.}, to be:
\beq \label{definition bgdist}d_{ns}(x,y):=\min d(w,z)\textrm{ where
} w\in J^-(x)\cap J^-(y) \textrm{ and }z\in J^+(x)\cap J^+(y) \;. \eeq It
is clear that for \emph{any} two unrelated elements $x,y$, the
minumum possible distance is $2$, when the relevant interval ($w,z$)
is empty, and we shall call the the \emph{trivial distance}. While the
definition of Eq.\ (\ref{definition bgdist}) works for causal sets
faithfully embedded into $\mathbf{M}^2$, it fails for higher
dimensions (as first noted in Ref.\ \cite{bg} and numerically
confirmed in Ref.\ \cite{diameters}). We will explore the reasons for
the failure so that we can then proceed to a new proposal that does
not suffer from those problems.

\subsection{Failure of \bgdist{}}

We first consider the continuum. In 1+1 dimensions, there exists a
unique pair of elements ($w,z$) from the common past to the common
future of the spacelike separated points in question ($x,y$), such
that the distance is minimum. The pair of points has $w$ at the
point of intersection of the past lightcones of $x$ and $y$, and
similarly $z$ the point of intersection of the %common
future lightcones.

In 2+1 dimensions, the locus of points of intersection of the two
past lightcones forms a hyperbola (a one dimensional object). For
\emph{each} of the points of the intersection of the past lightcones
($w$), there exist a unique point in the intersection of the future
lightcones ($z$), such that the timelike distance between the pair
$(w,z)$ is minimum (and thus equal to the spacelike distance between
our target points $x,y$). We name such pairs \emph{minimizing
pairs} and (in the continuum) there are clearly infinite of those.
Moreover, we can select a subset of those pairs (of infinite
cardinality), such that the intervals between any two of those pairs
has arbitrarily small overlap (we fix some small threshold
$\epsilon$). We name this collection of minimizing pairs
\emph{independent minimizing pairs} (IMP).

Now we return to the causal set that can be faithfully embedded in
2+1 dimensional Minkowski spacetime. It is not difficult to see that there will
be elements of the causal set close\footnote{The precise meaning of
``close'' is not important here, but to be accurate would require
more care.} to each of the IMP. Each of these pairs, gives (on
average) the correct distance, but since the causal set arises as a
poisson distribution, there is a finite (but very small) chance that
each pair will give a trivial distance. In other words there is a
finite probability for each $(w,z)$ IMP, that there will be no
elements between $w$ and $z$. Since we have infinite of those, we
are guaranteed that we will find \emph{at least} one IMP, that has
no elements between $w$ and $z$\footnote{The independent condition
is required because we need the $d(w,z)$ variables to be independent.
Otherwise, we would not be guaranteed to find one pair with zero
elements in between.}.

This is a very strong theoretical reason, spotted from the authors
of Ref.\ \cite{bg}. However, to check this on a computer is much more
difficult, because on a computer we cannot simulate a causal set faithfully
emebedded into
% do not actually have an
infinite
Minkowski spacetime, %but
only a finite region. The
\emph{independent} minimizing pairs are widely separated, %not too dense,
so testing
this effect is difficult, however in Ref.\ \cite{diameters} it was
clearly observed.

\section{Spatial Distance}\label{section spatial distance}

As we can see, the trouble with the above attempt was that we
minimized over all those pairs. Each of those had the correct
\emph{average} value, which means that taking an average over suitably
chosen pairs would solve this problem. We therefore need:

\begin{itemize}

%  Find the causet analogues of minimizing pairs,
% considering the causal set alone (i.e. without using the
% background).

\item[(a)] A mechanism to select causal set elements which lie close to a
  minimizing pair.  Such a mechanism involves:
\begin{itemize}
\item[(i)] Finding elements which are close to the intersection of the future
  (or past) light cones of our unrelated pair of elements.
\item[(ii)] For each such an element $z$ in the future (say), select an element
  $w$ in the common past which locates a pair $(w,z)$ which is close to some
  continuum minimizing pair.

\end{itemize}

\item[(b)]
To take an \textbf{average} over minimizing pairs, and not minimize
\end{itemize}
For the first issue, we require the definition of a \emph{2-link}:
Given unrelated elements $x,y$ of a causal set we define $w$ to be a
2-link (of $x,y$) if the element $w$ is linked to both $x$ and $y$
(see above for the definition of a link). Given an element $x$, the
elements that are linked to $x$ are elements that lie very close to
the lightcone of $x$. Therefore, it is easy to see that the
$2$-links lie close to the intersection of the lightcones of $x$ and
$y$. It is easy to see that for dimensions greater than 2, the
number of 2-links for any pair of unrelated elements $x,y$ is
infinite (corresponding to the (non-countably) infinite points in
the continuum). However in 1+1 this is not the case, and most
frequently there exist no 2-links at all. In what follows, we could
define things slightly differently to account for this problem in
1+1 dimensions\footnote{A concept of closest 2-neighbor that reduces
to a 2-link if such thing exists is one obvious attempt.}. However,
(a) we live in 3+1 dimensions and (b) in 1+1 dimensions the
\bgdist{} works fine, so we are not going to go into more details on
this issue here, and assume the existence of 2-links. We are now in
position to define the following procedure that gives us a spatial
distance that does not suffer from problems discussed earlier:

\begin{itemize}
\item[Step 1:] Given spacelike elements $x,y$ we find a future 2-link $f_i$.
\item[Step 2:] Find the element $p_i$ in the common past of $x$
and $y$ that makes the timelike distance $d(p_i,f_i)$ minimum.
This will select a minimizing pair.
\item[Step 3:] Store the timelike distance $d^i(x,y)$ for the future 2-link $f_i$.
\item[Step 4:] Repeat for all other future 2-links.
\item[Step 5:] Take the average over all future 2-links $\langle d^i(x,y) \rangle$ to be the spacelike
  distance between elements $x$ and $y$.  We call this average the
  \emph{2-link distance} between $x$ and $y$.
\end{itemize}
In Ref.\ \cite{diameters} we can see how this distance performs
numerically and compare it with \bgdist{} for both the cases that
\bgdist{} is valid and where it fails. The results of the
simulations agree with the intuitive theoretical expectations
analyzed above.

\section{Towards Curved Spacetime}\label{section curved spacetime}

\begin{figure}[hbtp]
\center\includegraphics{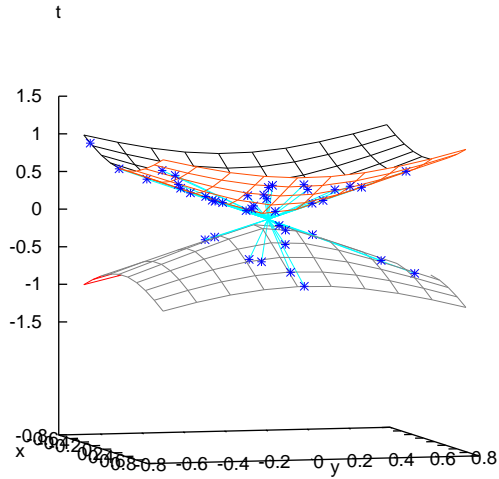}
\caption{\label{closest-neighbors}Spatial nearest neighbors of an
element in a sprinkling into (a
    fixed cube in) 2+1 dimensional owski.  $\langle N \rangle = 65\,536$.  The future and
    past light cones of the ``origin element'' $x$ are shown.  The spacelike
    cyan lines are drawn between $x$ and each neighbor, for emphasis.}
\end{figure}

In the previous section we considered only how to define spacelike
distance between elements of a causal set approximated by a
Minkowski spacetime. Using this, we can define a concept of closest
spatial neighbor. We define a (symmetric) relation \emph{s-link}
between to unrelated elements $x,y$. This relation exists if
the 2-link distance between
$x$ and $y$ is %are separated
less than some fixed threshold $\lambda$. This
threshold is somewhat arbitrary, but it should be between 2 and 3
(note that with the definition we have above, the smallest
conceivable spatial distance would be 2, however taking an average over
all the IMP would give a value greater than 2 value no matter how close in
the embedding the elements $x,y$ are). Given a target element at the
origin of 2+1 dimensional Minkowski spacetime, we can see in Figure
\ref{closest-neighbors}, that the s-links indeed lie on a
hyperboloid as one would expect.

Having defined the concept of an s-link, which is the most elementary
relation between unrelated (spacelike) elements, it is straight
forward to define what a continuous curve is: It is a collection of
elements that can be ordered in such a way that any two consecutive
elements are related either as links or as s-links. The length of
such continuous curve can simply be derived by counting the number
of links and s-links (or else the number of elements).

An important observation is that for spacetimes for which the curvature
does not vary rapidly at scales just above the Plank scale, the
prescription we define to identify closest neighbors would carry
over\footnote{It essentially says that for spacetimes that are
``locally flat'', in the sense defined above, we can identify which
are the closest neighbors using the prescription we mentioned above
for flat spacetimes.}. Identifying closest neighbors in \emph{any}
(flat or curved) spacetime means that we can compute lengths of
curves in those spacetimes. The very concept of spatial distance is
not well defined for curved spacetimes, however the length of curves
\emph{is} and the prescription we gave can achieve this.

Detailed numerical analysis, and possible extensions of these ideas
in order to recover the full metric, are left for further work.
Recovering the full metric may also help in giving direction for
deriving quantum dynamics of the causal set (e.g.\ by rewritting the
Einstein-Hilbert action in terms of the relations of the causal
set). However, already recovering the lengths of curves is an
important achievement in recovering the effective spacetime that
approximates our causal set and thus a major step forward in proving
the central conjecture.

\section{Summary and
Conclusions}\label{conclusions}

We have reviewed how to recover timelike distance from a causal set
and a failed attempt to do something similar for the spacelike
distance on a Minkowski causal set. The reason for this failure was
that there exist infinite independent ``minimizing pairs'' and the
definition of \bgdist{} involved minimizing over all of them. Since
they turn out to be random variables, minimizing over infinite of
those would give a trivial result. To avoid this, we suggested
taking an average over those minimizing pairs. However, to do so we
need to define what is precisely meant by a minimizing pair in a
causal set, which involved the concept of a 2-link. By considering,
the closest spatial neighbors, we were able to define length of
curves. Moreover, this relies only on closest neighbors, so for
causal sets that do not have curvature at (or close to) the Plank
scale, we can hope that the definition of closest neighbors carries
over intact. Therefore we can define length of curves in \emph{any}
causal set (flat or curved) that can be approximated by a continuum
spacetime. Further analysis of the curved spacetime case, and the
possibility of fully recovering the metric, is left for future work.

\ack We are especially grateful to Graham Brightwell for a number of
discussions on this work. We are also grateful to Rafael Sorkin,
David Meyer, and the attenders of `relativity lunch' at Imperial
College London, for numerous helpful discussions.

This research was supported by a number of grants/organizations,
including the Marie Curie Research and Training Network ENRAGE
(MRTN-CT-2004-005616), the Royal Society International Joint Project
2006-R2, and the Perimeter Institute for Theoretical Physics.
Research at Perimeter Institute is supported by the Government of
Canada through Industry Canada and by the Province of Ontario
through the Ministry of Research \& Innovation.

The numerical results were made possible by the facilities of the
Shared Hierarchical Academic Research Computing Network
(SHARCNET:www.sharcnet.ca).

PW thanks the Perimeter Institute for Theoretical Physics for
hospitality during a visit in which a large part of this work was carried
out, and the organizers of NEB XIII for giving the opportunity to
present this talk.

\end{document}